\documentclass[prb,twocolumn,showpacs,superscriptaddress,amsmath,amssymb]{revtex4}

\usepackage{graphicx}
\usepackage{latexsym}
\usepackage{amsmath}
\usepackage{amssymb}
\usepackage{amsfonts}
\usepackage{color}

\newcommand{\ds}{\displaystyle}

\begin{document}
\bibliographystyle{apsrev}

\title{Fluctuations in a mesoscopic superconducting ring: resonant behavior of conductivity and specific heat in two mode critical regime}

\author{S.V. Mironov}
\affiliation{Institute for Physics of Microstructures, Russian
Academy of Sciences, 603950 Nizhny Novgorod, GSP-105, Russia}
\author{A. Buzdin}
\affiliation{Institut Universitaire de France and University
Bordeaux, LOMA UMR-CNRS 5798, F-33405 Talence Cedex, France}

\date{\today}
\begin{abstract}
The critical fluctuations in a mesoscopic superconducting ring are
studied within the Ginzburg-Landau approach. The nonlocal
conductivity as well as the specific heat are calculated as
functions of the magnetic flux $\Phi$ through the ring. At
$\Phi=\Phi_0/2$ two low-energy eigenstates become degenerate and
near this point the behavior of fluctuations-dependent quantities
change dramatically: both the zero Fourier component $\sigma_0$ of
the fluctuation conductivity and the specific heat become
non-monotonic functions of $\left|\Phi-\Phi_0/2\right|$ with
rather special resonant structure.

\end{abstract}

\pacs{74.25.F-, 74.78.Na, 72.10.-d, 71.10.Ay, 71.30.+h.}

\maketitle

\section{Introduction}\label{Intro}

Mesoscopic superconducting rings attract considerable interest in
connection with their unusual properties in the vicinity of the
superconduction transition. One of the most interesting features
is an oscillatory behavior of the critical temperature as a
function of magnetic flux through the ring \cite{Little} due to
competition between spatial modes characterizing by different
orbital momenta. The remarkable property of the superconducting
rings is the fact that all eigenstates are well separated from
each other and in some cases the dynamics of each state can be
described independently. This provide a unique opportunity to
study the critical fluctuation contribution to all measurable
quantities of the rings. With the development of the
microfabrication technologies it becomes possible to create
superconducting rings of the radius $R$ comparable to the
Ginzburg-Landau coherence length $\xi_0$ (see
Refs.~\onlinecite{Koshnick, Morelle} and references therein). The
fluctuations in such small rings can be described in the frames of
simple analytical models, such as the zero-dimensional
Ginzburg-Landau formalism (see, for example,
Ref.~\onlinecite{Abrikosov}). In Refs.~\onlinecite{Schwiete_PRL,
Schwiete} the persistent current due to the critical fluctuations
was studied for different ratio between the rings' radius and the
coherence length. The calculation of magnetization in the critical
regime have been performed in Ref.~\onlinecite{Daumens}. Note that
previously the influence of the critical fluctuations on specific
heat and magnetization has been studied in small superconducting
granula both theoretically \cite{Schmidt} and experimentally
\cite{Buhrman}.

At the same time one of the most natural way to study fluctuations
near $T_c$ is the performance of the conductivity measurements
(the fluctuation correction to the conductivity has been described
first in Ref.~\onlinecite{Aslamazov}). For superconducting rings
two types of experiments were proposed: (i) nonlocal
paraconductivity measurements with probes attached to different
points of the ring \cite{Israeloff, Langer}; (ii) contactless
measurements of dissipation in the array of rings subject to an
electromotive force induced by a weak alternating magnetic field
\cite{Buzdin}. The corresponding calculations of the nonlocal
conductivity within the Gaussian approximation were performed in
Ref.~\onlinecite{Glazman}.

The separation of different eigenstates in mesoscopic ring with
$R\gtrsim \xi_0$ allows to describe analytically the contribution
of the critical fluctuations to the nonlocal conductivity in
magnetic field which produces the flux $\Phi$ through the ring. It
was shown that in the critical fluctuation regime the fluctuation
contribution to the conductivity has a logarithmic singularity
near the critical temperature $T_c$ \cite{Buzdin} with a prefactor
proportional to the $(\Phi/\Phi_0)^2$. However this approach is
valid only for magnetic flux values which are not close to
$\Phi_0/2$. Otherwise for $\Phi\approx\Phi_0/2$ the separation of
modes with different orbital momenta breaks down: critical
fluctuations near $T_c$ are produced by two interacting modes,
which can not be considered separately anymore.

In the present paper we suggest an analytical description of
critical fluctuations for the case $\Phi\approx\Phi_0/2$ and
calculate the corresponding fluctuation correction to the
non-local conductivity of the ring. We will keep in mind the
contactless realization of conductivity measurements in which only
the zero Fourier component of the conductivity plays the key role.
Also we analyze the behavior of the specific heat value in
two-mode critical regime. To analyze the situation we use
time-dependent Ginzburg-Landau equation. This approach is
reasonable for small dirty superconducting rings (see
Ref.~\onlinecite{Buzdin}).

The paper has the following structure. In Section \ref{Conduct} we
describe in detail the behavior of the conductivity both inside
and outside of the two-mode critical regime. In particular, we
present the exact expression for the conductivity for the case
$\phi\approx 1/2$, based on the analytical solution of nonlinear
two-mode Ginzburg-Landau equation. In Section \ref{SpecHeat} we
calculate the specific heat taking the mode interaction into
account. The results are summarized in Section \ref{Conclud}.

\section{Fluctuation conductivity due to critical fluctuations}\label{Conduct}

To calculate the non-local conductivity we use the approach which
is similar to that of Ref.~\onlinecite{Buzdin}. The zero-frequency
conductivity $\sigma(\varphi-\varphi')$ is given by the Kubo
formula (see, for example, Ref.~\onlinecite{Glazman})
\begin{equation}\label{Kubo}
\sigma(\varphi-\varphi')=\frac{1}{T}\int\limits_{0}^{\infty}
\left<J(\varphi,0)J(\varphi',t)\right>dt,
\end{equation}
where the supercurrent is defined by the standard expression
$J(\varphi,t)=(e/mR){\rm Re}\left[\psi\left(i\partial_{\varphi}-
\phi\right)\psi^{*}\right]$ and $\phi=\Phi/\Phi_0$.

Performing the Fourier transform of the order parameter wave
function and the conductivity
\begin{equation}\label{Sigma_Fourier_Transform}
\psi(\varphi,t)=\sum\limits_{n}\psi_n(t)
e^{in\varphi},\sigma(\varphi-\varphi')=\sum\limits_{k}\sigma_k
e^{ik(\varphi-\varphi')},
\end{equation}
one can obtain the following expression for the zero Fourier
component $\sigma_0$
\begin{equation}\label{Kubo_Fourier_0}
\begin{array}{r}{\ds
\sigma_0=\frac{4}{T}\left(\frac{e}{2mR}\right)^2\int\limits_{0}^{\infty}
dt\sum\limits_{n,m}\left(n-\phi\right)
\left(m-\phi\right)}\\{\ds\times
\left<|\psi_n(0)|^2|\psi_m(t)|^2\right>,}
\end{array}
\end{equation}
while for nonzero components $\sigma_{k\not= 0}$
\begin{equation}\label{Kubo_Fourier_k}
\begin{array}{r}
{\ds \sigma_k=\frac{1}{T}\left(\frac{e}{2mR}\right)^2
\int\limits_{0}^{\infty} dt\sum\limits_{n}\left(2n+k-2\phi\right)
}\\{\ds \times\left<\psi_n^{*}(0)\psi_n(t)
\psi_{n+k}(0)\psi_{n+k}^{*}(t)\right>}.
\end{array}
\end{equation}
The Fourier components of the order parameter wave function
$\psi_n(t)$ satisfy the non-linear time-dependent Ginzburg-Landau
equation \cite{Larkin}
\begin{equation}\label{GL_Harmonics_General}
-\gamma\partial_t\psi_n=
a(T)\psi_n+b\sum\limits_{k+m-p=n}\psi_k\psi_p^{*}\psi_m,
\end{equation}
where the values $a=(4m\xi_0^2)^{-1}[\varepsilon+
(\xi_0/R)^2\left(n-\phi\right)^2]$ and $b$ are the parameters of
the Ginzburg-Landau theory, $\varepsilon=(T-T_{c0})/T_{c0}$,
$\xi_0^{-2}=4m\alpha T_{c0}$ and $\gamma=\pi\alpha/8$.

Since the critical temperature $T_c(\phi)$ is a periodic function
with the period $1$ due to the Little-Parks effect and
$T_c(-\phi)=T_c(\phi)$ we can consider only the range
$0\leq\phi\leq 1/2$ for the magnetic flux. In this range
$T_c(\phi)=T_{c0}\left[1-(\xi_0/R)^2\phi^2\right]$ and it is
convenient to use the parameter
$\varepsilon^{*}=\left[T-T_c(\phi)\right]/T_{c0}=\varepsilon+(\xi_0/R)^2\phi^2$
instead of $\varepsilon$.

Let us analyze the dependencies of the conductivity on the
magnetic flux through the ring in different temperature ranges. In
case when when $\varepsilon^{*}\gg Gi_{(0)}$ the system is in the
Gaussian regime and one can neglect the nonlinearity in
Eq.~(\ref{GL_Harmonics_General}). Here we have introduced the
Ginzburg-Levanyuk number $Gi_{(0)}$ characterizing the width of
the critical fluctuation region in zero-dimensional systems:
$Gi_{(0)}=\sqrt{2b/\alpha^2T_cV}$, where $V=2\pi Rs$ is the volume
of the ring and $s$ is its cross-section. In the dirty limit a
simple estimate shows that $Gi_{(0)}\propto (T_c/E_F)[(\xi_0
l)^{3/4}/\sqrt{V}]$, where $l$ is the electron mean free path.
Overwise when $\varepsilon^{*}\ll Gi_{(0)}$ the fluctuations of
one or two lowest modes become critical and nonlinear terms in
Eq.~(\ref{GL_Harmonics_General}) begin to play the key-role in the
description of the fluctuation conductivity. Further we will
consider these fluctuation regimes separately.

\subsection{Gaussian regime}\label{Cond_Gauss}

In the Gaussian regime one can find from the linearized
Eq.~(\ref{GL_Harmonics_General}) that
$\psi_n(t)=\psi_n(0)\exp(-t/\tau_n)$, where
$\tau_n^{-1}=(8T_{c0}/\pi)\left[\varepsilon^{*}+(\xi_0/R)^2n(n-2\phi)\right]$.
Also it is easy to find the exact expressions for correlators
$\left<|\psi_n(0)|^2\right>$ and $\left<|\psi_n(0)|^4\right>$
which contribute to $\sigma_0$:
\begin{equation}\label{Corr_2}
\left<|\psi_n(0)|^2\right>=\frac{4m\xi_0^2T}{V
\left[\varepsilon^{*}+(\xi_0/R)^2n(n-2\phi)\right]},
\end{equation}
\begin{equation}\label{Corr_4}
\left<|\psi_n(0)|^4\right>=\frac{2(4m\xi_0^2)^2T^2}{V^2
\left[\varepsilon^{*}+(\xi_0/R)^2n(n-2\phi)\right]^2}.
\end{equation}

From the expression (\ref{Kubo_Fourier_0}) one can see that the
singular part of $\sigma_0$ comes only from terms with $m=0$ and
$m=1$. The corresponding expression for the fluctuation
conductivity in the Gaussian regime is rather complicated and
therefore is given in the Appendix. Here we will focus only on two
limiting cases.

For $\varepsilon^{*}\ll(\xi_0/R)^2(1-2\phi)$ the most singular
part in the temperature dependence of the conductivity comes from
the mode with $n=0$ and has the form
\begin{equation}\label{Sigma_Gauss_Far}
\sigma_0^{(G)}(\varepsilon^{*})=\frac{e^2}{4\pi
s^2}\left(\frac{\xi_0}{R}\right)^4\frac{\phi^2}{(\varepsilon^{*})^3}.
\end{equation}
For $\varepsilon^{*}\gg(\xi_0/R)^2(1-2\phi)$ in the vicinity of
the half quantum flux through the ring two modes with $n=0$ and
$n=1$ contribute to the most singular part of $\sigma_0$, which
reads as
\begin{equation}\label{Sigma_Gauss_Near}
\sigma_0^{(G)}(\varepsilon^{*})=\frac{e^2}{4\pi
s^2}\left(\frac{\xi_0}{R}\right)^4\frac{\phi^2+2(1-2\phi)^2}{(\varepsilon^{*})^3}.
\end{equation}
These expressions are valid for any $\phi$ from the range
$0\leq\phi\leq\ 1/2$.

\subsection{Critical regime with noninteracting modes}\label{Cond_Crit}

Let us first consider critical fluctuations for magnetic flux
values which are far from $\phi=1/2$. In this case only one
spatial mode with $m=0$ is in a critical regime and thus different
spatial modes do not interact with each other. Then the equation
(\ref{GL_Harmonics_General}) for $\psi_0(t)$ can be solved
analytically \cite{Buzdin}. Note that the dominant contribution to
the conductivity comes from the correlator with $n=m=0$ .
Considering only this correlator in the expression
(\ref{Kubo_Fourier_0}) (the lowest mode approximation) one can
reproduce the result of Ref.~\onlinecite{Buzdin} with corrected
numerical coefficient (the correlator
$\left<|\psi_0(0)|^2|\psi_0(t)|^2\right>$ should be calculated
instead of the correlator $\left<\psi_0(0)\psi^{*}_0(t)\right>^2$,
which provide an additional factor
$(8\pi)^{1/2}\Gamma^{-2}(1/4)\approx 0.38$):
\begin{equation}\label{Sigma_0_0}
\sigma_0^{(0)}(\phi)=\frac{e^2}{\pi^{3/2}s^2}\left(\frac{\xi_0}{R}\right)^4
\phi^2\frac{1}{Gi_{(0)}^3}{\rm ln}\frac{1}{\varepsilon^{*}}.
\end{equation}

Note that the lowest mode approximation, which has been used to
obtain the expression (\ref{Sigma_0_0}), is valid when the sum of
all neglected terms in (\ref{Kubo_Fourier_0}) is small compared
with $\sigma_0^{(0)}$. Let us calculate all terms with $m=0$ and
$m=1$ in the expression (\ref{Kubo_Fourier_0}) which can become
singular near $T_c(\phi)$ for $0\leq\phi\leq 1/2$. The
corresponding expression for the conductivity $\sigma_0^{(C)}$ is
given in the Appendix since it is rather cumbersome.

Near $\phi=1/2$ for $\varepsilon^{*}\ll(\xi_0/R)^2(1-2\phi)$ the
most singular correction to the conductivity (\ref{Sigma_0_0}) is
negative and can be written in the form
\begin{equation}\label{Sigma_0_correction}
\Delta\sigma_0^{(0)}=-\frac{e^2}{8\pi
s^2}\left(\frac{\xi_0}{R}\right)^2\frac{1}{Gi_{(0)}^2}\frac{1}{(1-2\phi)}{\rm
ln}\frac{1}{\varepsilon^{*}}.
\end{equation}

From the expression (\ref{Sigma_0_correction}) one can see that
for $\left|\phi-1/2\right|< \left(R/\xi_0\right)^2 Gi_{(0)}$ the
value $\Delta\sigma_0^{(0)}/\sigma_0^{(0)}$ exceeds 1 which
indicates that the expression (\ref{Sigma_0_0}) is not valid in
this region. Note that in our case $R\gtrsim \xi_0$ and the
typical value for the Ginzburg-Levanyuk number is $Gi_{(0)}\sim
10^{-3}$ (for Al rings \cite{Israeloff}), so the lowest mode
approximation breaks down only in a narrow region near the point
$\phi=1/2$ when both modes with $m=0$ and $m=1$ are strongly
fluctuating and interact with each other. The case when
$\left|\phi-1/2\right|< \left(R/\xi_0\right)^2 Gi_{(0)}$ is
considered in the next subsection.

\subsection{Two-mode critical regime}

To obtain the expression for the critical fluctuations'
contribution to paraconductivity at the magnetic flux $\phi\approx
1/2$ one should consider the dynamics of two lowest interacting
modes. Further we restrict ourselves to the most interesting case
when the magnetic flux through the ring is exactly half quantum
$\phi=1/2$. In this case all modes except the modes with $n=0,1$
can be neglected since they do not contribute to the singular part
of $\sigma_0$. Then the expression for $\sigma_0$ reads
\begin{equation*}\label{Kubo_Fourier_0_2modes}
\begin{array}{l}{\ds
\sigma_0=\frac{1}{T}\left(\frac{e}{2mR}\right)^2}\\{\ds\times
\left<\left(|\psi_0(0)|^2-|\psi_1(0)|^2\right)\int\limits_{0}^{\infty}
\left(|\psi_0(t)|^2-|\psi_1(t)|^2\right)dt\right>.}
\end{array}
\end{equation*}

The system of nonlinear time-dependent Ginzburg-Landau equations
for $\psi_0(t)$ and $\psi_1(t)$ is
\begin{equation}\label{System_start}
\left\{
\begin{array}{c}
{\ds \gamma\partial_t\psi_0 +
\frac{\varepsilon^{*}}{4m\xi_0^2}\psi_0
+b\left(|\psi_0|^2+2|\psi_1|^2\right)\psi_0=0,}\\{\ds
\gamma\partial_t\psi_1 + \frac{\varepsilon^{*}}{4m\xi_0^2}\psi_1
+b\left(|\psi_1|^2+2|\psi_0|^2\right)\psi_1=0,}
\end{array}
\right.
\end{equation}
where $\varepsilon^{*}=\varepsilon+\xi_0^2/4R^2$. Let us multiply
the first equation of this system by $\psi_0^{*}$, the second
equation by $\psi_1^{*}$ and then sum each of the obtained
equations with its complex conjugated. The result is
\begin{equation}\label{System_modules}
\left\{
\begin{array}{c}
{\ds \partial_{\tilde{t}} F + a^{*}F+F^2+2FG=0,}\\
{\ds \partial_{\tilde{t}} G+ a^{*}G+G^2+2FG=0,}
\end{array}
\right.
\end{equation}
where $\tilde{t}=16bt/\pi\alpha$, $a^{*}=\varepsilon^{*}/4m\xi_0^2
b$, $F=|\psi_0|^2$ and $G=|\psi_1|^2$. To simplify these equations
we introduce a new time variable $\tau=(a^{*})^{-1}\left[1-{\rm
exp}(-a^{*}\tilde{t})\right]$ and new functions
$f(\tau)=e^{a^{*}\tilde{t}(\tau)}F(\tilde{t}(\tau))$ and
$g(\tau)=e^{a^{*}\tilde{t}(\tau)}G(\tilde{t}(\tau))$. Then the
system (\ref{System_modules}) transforms into the form
\begin{equation}\label{System_NewTime}
\left\{
\begin{array}{c}
{\ds \partial_{\tau} f +f^2+2fg=0,}\\
{\ds \partial_{\tau} g+g^2+2fg=0.}
\end{array}
\right.
\end{equation}

To obtain the exact solution of the system (\ref{System_NewTime})
it is convenient to consider the auxiliary functions $u=(f+g)/2$
and $v=(f-g)/2$. Note that $\sigma_0$ depends only on $v(\tau)$:
\begin{equation}\label{Kubo_Fourier_0_2modes_u}
\sigma_0=\frac{1}{T}\left(\frac{e}{mR}\right)^2
\frac{\pi\alpha}{16b}\left<v(0)\int\limits_{0}^{1/a^{*}}v(\tau)d\tau\right>.
\end{equation}
From the system (\ref{System_NewTime}) we obtain the equations for
new functions
\begin{equation}\label{uvw_eq}
\partial_{\tau} u=v^2-3u^2,~~~~~~\partial_{\tau} v=-2uv,
\end{equation}
which can be reduced to the equation for the $v$ function
\begin{equation}\label{v_eq}
2v\partial^2_{\tau}v-5\left(\partial_{\tau}v\right)^2+4v^4=0.
\end{equation}
Then for the decaying solution of Eq.~(\ref{v_eq}) we find
\begin{equation}\label{v_eq_mod}
\partial_{\tau}v=-2\mu v^2\sqrt{1+\lambda v},
\end{equation}
where $\lambda=(u_0^2-v_0^2)/v_0^3$ (here and after the index 0
indicates the function value at $\tau=0$ and $v_0$ is assumed to
be nonzero), $\mu={\rm sign}(v_0)$.

Note that the explicit solution of the system
(\ref{System_NewTime}) integrated over time has the form
\begin{equation}\label{Solution}
\begin{array}{l}{\ds
\int\limits_0^{\infty}f(t)dt=\left\{\begin{array}{l} {\ds {\rm
ln}\frac{2(f_0-g_0)^2}{f_0 a^{*}}~~~ {\rm for}~~ \mu=1,}\\{}\\{\ds
{\rm ln}\frac{g_0}{\left|f_0-g_0\right|}~~~~~~ {\rm for}~~
\mu=-1;}\end{array}\right.}\\{}\\{\ds
\int\limits_0^{\infty}g(t)dt=\left\{\begin{array}{l} {\ds {\rm
ln}\frac{f_0}{\left|f_0-g_0\right|}~~~~~~ {\rm for}~~
\mu=1;}\\{}\\{\ds {\rm ln}\frac{2(f_0-g_0)^2}{g_0 a^{*}}~~~ {\rm
for}~~ \mu=-1.}\end{array}\right.}
\end{array}
\end{equation}
An interesting feature of the solution (\ref{Solution}) is that
depending on the initial condition only one mode ($f$ or $g$)
becomes slowly decaying due to the nonlinear modes' interaction.

\begin{figure}[t!]
\includegraphics[width=0.5\textwidth]{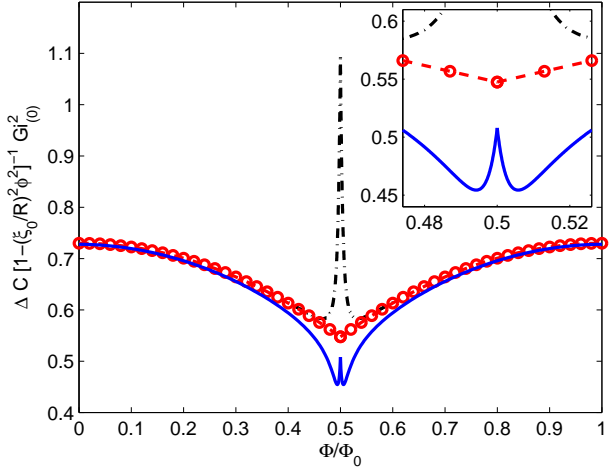}
\caption{(color online) The specific heat as a function of
magnetic flux $\phi$ through the ring at the temperature of the
superconducting transition for $\left(R/\xi_0\right)^2
Gi_{(0)}=10^{-2}$. The exact solution (blue solid curve) near
$\phi=1/2$ qualitatively differs from the solution obtained taking
only zero spatial mode into account (red dashed curve with
circles) and is in sharp contrast with the predictions of the
independent modes approximation (black dashdot curve).}
\label{Fig_C_Phi}
\end{figure}

The expression (\ref{v_eq_mod}) allows us to calculate $\sigma_0$
since
\begin{equation}\label{Integral}
\int\limits_{0}^{1/a^{*}}v(\tau)d\tau=\int\limits_{v_0}^{v(1/a^{*})}\frac{v}{\left(\partial_{\tau}
v\right)}dv.
\end{equation}
To estimate the $v(1/a^{*})$ value at the temperature of the
superconducting transition ($a^{*}\to 0$) we use the asymptotic
form of the equation (\ref{v_eq_mod}) solution, which gives us
$v\left(1/a^{*}\right)\approx \mu a^{*}/2$. Then the singular part
of the integral (\ref{v_eq_mod}) has the form
\begin{equation}\label{Integral_Result}
\int\limits_{0}^{1/a^{*}}v(\tau)d\tau\approx\frac{|v_0|}{2}{\rm
ln}\frac{1}{\varepsilon^{*}}.
\end{equation}

\begin{figure}[t!]
\includegraphics[width=0.5\textwidth]{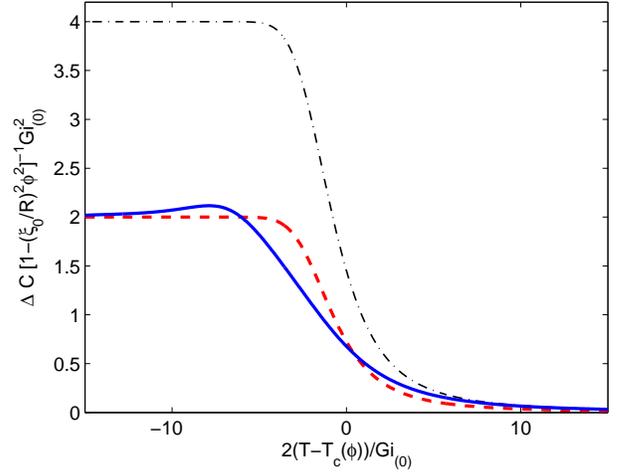}
\caption{(color online) The dependencies of the specific heat on
temperature for $\phi=0$ (red dashed curve) and $\phi=1/2$ (blue
solid curve). The black dashdot curve corresponds to the
prediction of the independent modes approximation.}
\label{Fig_C_T}
\end{figure}

The Gibbs free energy functional for two considered modes has the
form
\begin{equation}\label{Gibbs_energy}
F=V\left[a^{*}\left(f_0+g_0\right)+\frac{b}{2}\left(f_0^2+g_0^2+4f_0g_0\right)\right].
\end{equation}
Performing the integration over $f_0$ and $g_0$ in the expression
(\ref{Kubo_Fourier_0_2modes_u}) in the limit $a^{*}\to 0$ we
obtain
\begin{equation}\label{Sigma_0_answer}
\sigma_0(\phi=1/2)=
\Lambda\frac{e^2}{\pi^{3/2}s^2}\left(\frac{\xi_0}{R}\right)^4
\left(\frac{1}{2}\right)^2\frac{1}{Gi_{(0)}^3}{\rm
ln}\frac{1}{\varepsilon^{*}},
\end{equation}
where $\Lambda=\pi(\sqrt{3}-1)\left[\sqrt{2}{\rm
arcth}\left(\sqrt{2/3}\right)\right]^{-1}\approx 1.42$. Thus at
$\phi=1/2$ the exact expression of $\sigma_0$ is 1.42 times larger
than the one obtained within the lowest mode approximation (see
the expression (\ref{Sigma_0_0})).

Note that for $\phi=1/2$ the nonzero components $\sigma_k$ do not
diverge at $\varepsilon^{*}=0$. Indeed only the correlators in
(\ref{Kubo_Fourier_k}) that contain $\psi_0$ or $\psi_1$ functions
can produce the singularity of $\sigma_k$ at $\varepsilon^{*}=0$.
But the prefactor of the corresponding correlator in the
$\sigma_1$ expression is zero in case $\phi=1/2$ and thus the
$\sigma_1$ component does not contain the singular part. To
calculate other $\sigma_k$ components we should take into account
the following: (i) the correlators which contain noncritical modes
can be represented as a product of two independent corelators;
(ii) the corelators for critical modes can be calculated only for
$\tau=0$ since near $\varepsilon^{*}=0$ the decay times $t_n=\pi
R^2/8T\xi_0^2n(n-1)$ for noncritical modes are much shorter than
for modes with $n=0,1$; (iii) for noncritical modes
$\left<\psi_n(0)\psi_n^{*}(t)\right>=\left<|\psi_n(0)|^2
\right>t_n$. Note also that for $\phi=1/2$ the correlators
$\left<f_0\right>=\left<g_0\right>$. Therefore one can easily make
sure that $\sigma_k$ does not contain a singular part at all since
all contributions from critical modes cancel. Thus at $\phi=1/2$
only the $\sigma_0$ component has a singularity at
$\varepsilon^{*}=0$.

It is interesting to analyze the conductivity behavior when the
flux is slightly smaller than half-quantum $\delta
\phi=1/2-\phi\ll 1$. Then the system of equation
(\ref{System_NewTime}) has the form
\begin{equation}\label{System_NewTime_Delta}
\left\{
\begin{array}{l}
{\ds \partial_{\tau} f +f^2+2fg=0,}\\ {\ds \partial_{\tau}
g+\frac{\delta a^{*}}{(1-a^{*}\tau)}g+g^2+2fg=0,}
\end{array}
\right.
\end{equation}
where $\delta a^{*}=\delta \phi/(2mR^2b)$. Note that here the
value $a^{*}$ which is used in the definition of functions $f$ and
$g$ corresponds to modified critical temperature and reads as
$a^{*}=(4m\xi_0^2)^{-1}
\left[\varepsilon+(\xi_0/R)^2(1/2-\delta\phi)^2\right]$. In what
follows we will assume that $\delta \phi\ll \left(R/\xi_0\right)^2
Gi_{(0)}$.

For $f_0>g_0$ the dynamics of the slow-decaying mode $f$, which
define the singular contribution to the conductivity, is weakly
affected by the differences in the character of the mode $g$
decaying due to finite $\delta a^{*}$. At the same time for
$g_0>f_0$ the situation changes. For the $\tau$ values when
$g>>\delta a^{*}/(1-a^{*}\tau)$ the mode $g$ is slowly decaying
while at $\tau\sim (a^{*}+\delta a^{*})^{-1}$ which corresponds to
the case $g\sim\delta a^{*}/(1-a^{*}\tau)$ the decaying of the
$g$-mode becomes exponential. Thus for $g_0>f_0$ one has to
replace the cutoff $1/a^{*}$ in the integral
(\ref{Integral_Result}) with $1/(a^{*}+\delta a^{*})$. Then the
expression (\ref{Sigma_0_answer}) becomes modified as follows
\begin{equation}\label{Sigma_0_answer_Delta}
\sigma_0(\delta\phi)= \frac{\Lambda
e^2}{8\pi^{3/2}s^2}\left(\frac{\xi_0}{R}\right)^4
\frac{1}{Gi_{(0)}^3}\left({\rm ln}\frac{1}{\varepsilon^{*}}+{\rm
ln}\frac{1}{\varepsilon^{*}+\delta\varepsilon^{*}}\right),
\end{equation}
where $\delta\varepsilon^{*}=2(\xi_0/R)^2\delta \phi$.

It is interesting to compare the expressions
(\ref{Sigma_0_correction}), (\ref{Sigma_0_answer}) and
(\ref{Sigma_0_answer_Delta}). One can see that at magnetic fluxes
close to $\phi=1/2$ the conductivity has rather special resonant
structure. Indeed far from $\phi=1/2$ the conductivity is well
described with the expression (\ref{Sigma_0_0}). When $\phi$
increases the conductivity becomes smaller compared with the
expression (\ref{Sigma_0_0}) since the mode with $m=1$ begins to
enter the critical regime. This picture is broken down when
$1/2-\phi<<(R/\xi_0)^2Gi_{(0)}$, where the logarithmic singularity
in conductivity $\sigma_0(\phi)$ reveals a very sharp peak due to
the nonlinear modes' interaction. The maximum of the peak at
$\phi=1/2$ is larger than the one obtained in the lowest mode
approximation (see the expression (\ref{Sigma_0_0})). The width of
the peak is decreasing while $\varepsilon^{*}\to 0$ (see the
expression (\ref{Sigma_0_answer_Delta})).

\section{Fluctuation specific heat}\label{SpecHeat}

Now let us turn to the calculation of the fluctuation specific
heat $\Delta C(\phi,T)$. The explicit form of the Gibbs energy
$G[\psi]$ for two interacting modes with $n=0,1$ was obtained in
Ref.~\onlinecite{Daumens}. Then the specific heat can be
calculated as $\Delta C=-T\partial^2_T G$. For
$\left|\phi-1/2\right|> \left(R/\xi_0\right)^2 Gi_{(0)}$ the main
contribution to the $\Delta C$ at $\varepsilon^{*}\to 0$ comes
from the mode with $n=0$, which can be considered independently
from other modes. Then
\begin{equation}\label{C_OneMode}
\Delta
C(T=T_{c}(\phi))=\frac{2(\pi-2)}{\pi}\frac{1}{Gi_{(0)}^2}\left(1-\frac{\xi_0^2}{R^2}\phi^2\right).
\end{equation}
The corresponding dependence $\Delta C(\phi)$ at
$\varepsilon^{*}=0$ is shown in Fig.~1 (red dashed curve with
circles). Note that at $\phi=1/2$ the exact calculation gives the
value $\Delta C=\Theta~Gi_{(0)}^{-2}[1-(\xi_0/2R)^2]$, where
$\Theta=(4/3)(1+\sqrt{3}\eta-\pi\eta^2)$, $\eta=\left({\rm
arcch}~2\right)^{-1}$.

Moreover it is interesting to analyze the dependence of the
specific heat on temperature. In Fig.~2 two typical dependencies
$\Delta C(T)$ are shown for $\phi=0$ (red dashed curve) and for
$\phi=1/2$ (blue solid curve). In addition to the differences
between plotted curves near the critical temperature one can see
that in the superconducting state at $\phi=1/2$ the $\Delta C$ has
shallow peak due to the interaction of two critical modes.

Note that the exact solution for the specific heat in two-modes
critical regime is in a sharp contrast with the predictions
obtained within the independent modes approximation. For
comparison in Fig.~1 and 2 we have plotted corresponding
dependencies calculated without considering the modes' interaction
(black dashdot curves). One can see that these curves differ
qualitatively from the ones corresponding to exact solutions.

\section{Conclusion}\label{Conclud}

Thus we described the critical fluctuations in the regime when two
interacting modes are strongly fluctuating. Both specific heat and
magnetoconductivity are reveal special resonant behavior for
magnetic flux $\phi\approx 1/2$, which may serve as clear
indication of the two-mode critical regime for eventual
experiments. Note that our results obtained for superconducting
ring can be generalized directly for superconducting disk
\cite{Buisson} or the superconducting layer with the hole of the
round shape \cite{Bezryadin}. In these systems the magnetic field
which exceeds the upper critical field $H_{c2}$ but is less than
the field of the surface superconductivity $H_{c3}$ leads to
appearance of the superconducting nucleus localized in the ring
with the width of the order of $\xi_0$ near the superconductor
edge.

The predicted two modes critical fluctuations regime produces a
relatively strong variation of magnetoconductivity and specific
heat (of the order of 30\%-100\% near the half flux field). The
first detailed measurements of the paraconductivity in small
superconducting loops have been reported already in
Ref.~\onlinecite{Israeloff}, and recently the precise contactless
studies of the magnetic response in individual normal metal rings
have been performed \cite{Bluhm}. Therefore the experimental
studies of the critical regime in magnetoconductivty seem to be
quite feasible. Five years ago the sensitive attojoule calorimetry
\cite{Bourgeois, Ong} revealed the specific heat oscillations in
magnetic field for large arrays (450 thousands) of noninteracting
micrometer-sized superconducting loops. In such experiments the
averaging effects may play an important role. Unfortunately the
statistical characteristics of the studied ensemble of loops are
unknown, which does not allow comparing their results with our
calculations. However the recent progress in this domain should
permit to perform the calorimetric measurements on a single
superconducting ring in near future.

\section*{ACKNOWLEDGEMENTS}\label{Acknow}

The authors thank M. Daumens, A. Varlamov and A. Mel'nikov for
many useful discussions and suggestions. This work was supported
by the European IRSES program SIMTEC, French ANR ''SINUS", the
RFBR, Presidential RSS Council (Grant No.MK-4211.2011.2), RAS
under the Program ``Quantum physics of condensed matter", the
``Dynasty'' Foundation and FTP ``Scientific and educational
personnel of innovative Russia in 2009--2013".

\begin{widetext}

\appendix

\section*{APPENDIX}

In the Gaussian regime one can calculate the part of the
conductivity which can become singular near the temperature of the
superconducting transition for $0\leq\phi\leq 1/2$, considering
only terms with $m=0$ and $m=1$ in the expression
(\ref{Kubo_Fourier_0}). Then using the fact that in the Gaussian
regime for $k\not=l$ the identity
$\left<|\psi_k(0)|^2|\psi_l(0)|^2\right>=\left<|\psi_k(0)|^2
\right>\left<|\psi_l(0)|^2\right>$ is true and that $\int
\limits_0^{\infty}{\rm exp}\left(-t/\tau_n\right)dt=\tau_n$ we
obtain

\begin{equation}\label{Kubo_Fourier_0_explicit_1}
\begin{array}{r}{\ds
\sigma_0=\frac{1}{T}\left(\frac{e}{mR}\right)^2\left[\phi^2\tau_0
\left<|\psi_0(0)|^4\right>-\phi\tau_0\left<|\psi_0(0)|^2\right>
\sum\limits_{n\not=0 }\left(n-\phi\right)
\left<|\psi_n(0)|^2\right>\right.}\\{\ds \left.
+\left(1-\phi\right)^2\tau_1
\left<|\psi_1(0)|^4\right>+\left(1-\phi\right)\tau_1\left<|\psi_1(0)|^2\right>
\sum\limits_{n\not=1 }\left(n-\phi\right)
\left<|\psi_n(0)|^2\right> \right].}
\end{array}
\end{equation}
Substituting the expressions (\ref{Corr_2}) and (\ref{Corr_4})
into (\ref{Kubo_Fourier_0_explicit_1}) we obtain

\begin{equation}\label{Kubo_Fourier_0_explicit_2}
\begin{array}{r}{\ds
\sigma_0=\frac{1}{T}\left(\frac{e}{mR}\right)^2\frac{\left(4m\xi_0^2T\right)^2}{V^2}\left[
\frac{2\phi^2\tau_0}{\left(\varepsilon^{*}\right)^2}-\frac{\phi\tau_0}{\varepsilon^{*}}
\sum\limits_{n\not=0 } \frac{\left(n-\phi\right)}{
\left[\varepsilon^{*}+(\xi_0/R)^2n(n-2\phi)\right]}\right.}\\{\ds
\left. + \frac{2\left(1-\phi\right)^2\tau_1}{
\left[\varepsilon^{*}+(\xi_0/R)^2(1-2\phi)\right]^2}+\frac{\left(1-\phi\right)\tau_1}{
\left[\varepsilon^{*}+(\xi_0/R)^2(1-2\phi)\right]}\sum\limits_{n\not=1
} \frac{\left(n-\phi\right)}{
\left[\varepsilon^{*}+(\xi_0/R)^2n(n-2\phi)\right]} \right].}
\end{array}
\end{equation}
To calculate the principal values of the sums in the expression
(\ref{Kubo_Fourier_0_explicit_2}) near the critical temperature
one can put $\varepsilon^{*}=0$ in all terms except those with
$n=0,1$. The results are

\begin{equation}\label{Sum_1}
\begin{array}{c}{\ds \sum\limits_{n\not=0 } \frac{\left(n-\phi\right)}{
\left[\varepsilon^{*}+(\xi_0/R)^2n(n-2\phi)\right]}=}\\{=\ds
\frac{\left(1-\phi\right)}{
\varepsilon^{*}+(\xi_0/R)^2(1-2\phi)}-\frac{\left(1+\phi\right)}{
(\xi_0/R)^2(1+2\phi)}+\sum\limits_{n\geq 2}\left[
\frac{\left(n-\phi\right)}{
(\xi_0/R)^2n(n-2\phi)}-\frac{\left(n+\phi\right)}{
(\xi_0/R)^2n(n+2\phi)}\right]=}\\{\ds =\ds
\frac{\left(1-\phi\right)}{
\varepsilon^{*}+(\xi_0/R)^2(1-2\phi)}-\frac{\left(1+\phi\right)}{
(\xi_0/R)^2(1+2\phi)}+2\phi\left(\frac{R}{\xi_0}\right)^2
\left[\frac{1-12\phi^2}{8\phi^2\left(1-4\phi^2\right)}-\frac{\pi}{4\phi}{\rm
ctg}\left(2\pi\phi\right)\right]}
\end{array}
\end{equation}
and
\begin{equation}\label{Sum_1}
\sum\limits_{n\not=1} \frac{\left(n-\phi\right)}{
\left[\varepsilon^{*}+(\xi_0/R)^2n(n-2\phi)\right]}= -\frac{\phi}{
\varepsilon^{*}}-\frac{\left(1+\phi\right)}{
(\xi_0/R)^2(1+2\phi)}+\left(\frac{R}{\xi_0}\right)^2
\left[\frac{1-12\phi^2}{4\phi\left(1-4\phi^2\right)}-\frac{\pi}{2}{\rm
ctg}\left(2\pi\phi\right)\right].
\end{equation}
Summarizing we obtain the final result for the singular part of
the component $\sigma_0^{(G)}$ in the Gaussian regime:

\begin{equation}\label{Sigma_Gauss}
\begin{array}{c}{\ds
\sigma_0^{(G)}(\phi,\varepsilon^{*})=\frac{e^2}{4\pi
s^2}\left(\frac{\xi_0}{R}\right)^4\frac{\phi^2}{(\varepsilon^{*})^3}+\frac{e^2}{2\pi
s^2}\left(\frac{\xi_0}{R}\right)^4\frac{(1-\phi)^2}{\left[\varepsilon^{*}+(\xi_0/R)^2(1-2\phi)\right]^3}-}\\{\ds
- \frac{e^2}{4\pi
s^2}\left(\frac{\xi_0}{R}\right)^2\frac{\phi}{(\varepsilon^{*})^2}\left[\left(\frac{\xi_0}{R}\right)^2\frac{(1-\phi)}{
\left[\varepsilon^{*}+(\xi_0/R)^2(1-2\phi)\right]}-\frac{(1+\phi)}{(1+2\phi)}
+\left(\frac{1-12\phi^2}{4\phi(1-4\phi^2)}- \frac{\pi}{2}{\rm
ctg}(2\pi\phi)\right)\right]+}\\{\ds +\frac{e^2}{4\pi
s^2}\left(\frac{\xi_0}{R}\right)^2\frac{(1-\phi)}{\left[\varepsilon^{*}+(\xi_0/R)^2(1-2\phi)\right]^2}
\left[-\left(\frac{\xi_0}{R}\right)^2\frac{\phi}{
\varepsilon^{*}}-\frac{(1+\phi)}{(1+2\phi)}
+\left(\frac{1-12\phi^2}{4\phi(1-4\phi^2)}- \frac{\pi}{2}{\rm
ctg}(2\pi\phi)\right)\right].}
\end{array}
\end{equation}

For magnetic flux values which are far from the point $\phi=1/2$
the analogous singular part of the conductivity in the critical
regime can be calculated under the assumption that the
fluctuations of only the mode with $m=0$ are critical while the
fluctuations of other spatial modes are Gaussian (the lowest mode
approximation). The procedure of calculation is similar to the one
for the Gaussian regime. The expression (\ref{Kubo_Fourier_0})
takes the form

\begin{equation}\label{Kubo_Fourier_0_crit_1}
\begin{array}{r}{\ds
\sigma_0=\frac{1}{T}\left(\frac{e}{mR}\right)^2\left[\phi^2
\left<|\psi_0(0)|^2\int\limits_0^{\infty}
|\psi_0(t)|^2dt\right>-\phi\left<\int\limits_0^{\infty}
|\psi_0(t)|^2dt\right>\sum\limits_{n\not=0 }\left(n-\phi\right)
\left<|\psi_n(0)|^2\right>\right.}\\{\ds \left.
+\left(1-\phi\right)^2\tau_1
\left<|\psi_1(0)|^4\right>+\left(1-\phi\right)\tau_1
\left<|\psi_1(0)|^2\right>\sum\limits_{n\not=1
}\left(n-\phi\right) \left<|\psi_n(0)|^2\right> \right].}
\end{array}
\end{equation}
The first term in the square brackets in the expression
(\ref{Kubo_Fourier_0_crit_1}) represents the central result of
Ref.~\onlinecite{Buzdin}. To calculate the second term one should
take into account that in the critical regime

\begin{equation}\label{Corr_crit_2}
\left<\int\limits_0^{\infty} |\psi_0(t)|^2dt\right>=\frac{m
\xi_0^2}{4 Rs}\frac{1}{Gi_{(0)}^2}{\rm
ln}\frac{1}{\varepsilon^{*}}.
\end{equation}
The last sum in the square brackets in the expression
(\ref{Kubo_Fourier_0_crit_1}) is similar to the one in the
expression (\ref{Kubo_Fourier_0_explicit_1}) except the term with
$n=0$, which should be calculated in the critical regime. The
expression for the corresponding correlator in the explicit form
is given in Ref.~\onlinecite{Buzdin}. Summarizing we obtain the
resulting expression for the conductivity in the critical regime
for magnetic flux values which are far from $\phi=1/2$:

\begin{equation}\label{Sigma_Full}
\begin{array}{c}{\ds
\sigma_0^{(C)}(\phi)=\frac{e^2}{\pi^{3/2}s^2}\left(\frac{\xi_0}{R}\right)^4
\phi^2\frac{1}{Gi_{(0)}^3}{\rm
ln}\frac{1}{\varepsilon^{*}}+\frac{e^2}{2\pi
s^2}\left(\frac{\xi_0}{R}\right)^4
\frac{(1-\phi)^2}{\left[\varepsilon^{*}+\left(\xi_0/R\right)^2
\left(1-2\phi\right)\right]^3}-}\\{\ds-\frac{e^2}{2\pi
s^2}\phi\left(\frac{\xi_0}{R}\right)^2\frac{1}{Gi_{(0)}^2}{\rm
ln}\frac{1}{\varepsilon^{*}}
\left[\left(\frac{\xi_0}{R}\right)^2\frac{(1-\phi)}{
\left[\varepsilon^{*}+(\xi_0/R)^2(1-2\phi)\right]}-\frac{(1+\phi)}{(1+2\phi)}
+\left(\frac{1-12\phi^2}{4\phi(1-4\phi^2)}- \frac{\pi}{2}{\rm
ctg}(2\pi\phi)\right)\right]-}\\{}\\{\ds -\frac{e^2}{2\pi^{3/2}
s^2} \left(\frac{\xi_0}{R}\right)^6 \frac{1}{Gi_{(0)}}
\frac{\phi\left(1-\phi\right)}{\left[\varepsilon^{*}+\left(\xi_0/R\right)^2
\left(1-2\phi\right)\right]^2}.}
\end{array}
\end{equation}

Thus the expressions (\ref{Sigma_Gauss}) and (\ref{Sigma_Full})
give the singular part of the conductivity $\sigma_0$ in the
Gaussian and critical (in the lowest mode approximation) regimes
respectively.

\end{widetext}

\end{document}